\documentclass[sigconf, nonacm]{acmart}

\usepackage[ruled,vlined]{algorithm2e}
\usepackage{textcomp}
\newcommand\vldbdoi{XX.XX/XXX.XX}
\newcommand\vldbpages{XXX-XXX}
\newcommand\vldbvolume{14}
\newcommand\vldbissue{1}
\newcommand\vldbyear{2020}
\newcommand\vldbauthors{\authors}
\newcommand\vldbtitle{\shorttitle} 
\newcommand\vldbavailabilityurl{http://vldb.org/pvldb/format_vol14.html}
\newcommand\vldbpagestyle{empty} 

\begin{document}
\title{Protecting Big Data Privacy Using Randomized Tensor Network Decomposition and Dispersed Tensor Computation [Experiments, Analyses \& Benchmarks]}

\author{Jenn-Bing Ong}
\affiliation{%
  \institution{Nanyang Technological University}
  \country{Singapore}
}
\email{ongj0063@e.ntu.edu.sg}

\author{Wee-Keong Ng}
\affiliation{%
  \institution{Nanyang Technological University}
  \country{Singapore}
}

\author{Ivan Tjuawinata}
\affiliation{%
  \institution{Nanyang Technological University}
  \country{Singapore}
}

\author{Chao Li}
\affiliation{%
  \institution{RIKEN, Tokyo}
  \country{Japan}
}

\author{Jielin Yang}
\affiliation{%
  \institution{Nanyang Technological University}
  \country{Singapore}
}

\author{Sai None Myne}
\affiliation{%
  \institution{Singapore Management University}
  \country{Singapore}
}

\author{Huaxiong Wang}
\affiliation{%
  \institution{Nanyang Technological University}
  \country{Singapore}
}

\author{Kwok-Yan Lam}
\affiliation{%
  \institution{Nanyang Technological University}
  \country{Singapore}
}

\author{C.-C. Jay Kuo}
\affiliation{%
  \institution{University of Southern California}
  \country{United States of America}
}

\begin{abstract}
Data privacy is an important issue for organizations and enterprises to securely outsource data storage, sharing, and computation on clouds / fogs. However, data encryption is complicated in terms of the key management and distribution; existing secure computation techniques are expensive in terms of computational / communication cost and therefore do not scale to big data computation. Tensor network decomposition and distributed tensor computation have been widely used in signal processing and machine learning for dimensionality reduction and large-scale optimization. However, the potential of distributed tensor networks for big data privacy preservation have not been considered before, this motivates the current study. Our primary intuition is that tensor network representations are mathematically non-unique, unlinkable, and uninterpretable; tensor network representations naturally support a range of multilinear operations for compressed and distributed / dispersed computation. Therefore, we propose randomized algorithms to decompose big data into randomized tensor network representations and analyze the privacy leakage for 1D to 3D data tensors. The randomness mainly comes from the complex structural information commonly found in big data; randomization is based on controlled perturbation applied to the tensor blocks prior to decomposition. The distributed tensor representations are dispersed on multiple clouds / fogs or servers / devices with metadata privacy, this provides both distributed trust and management to seamlessly secure big data storage, communication, sharing, and computation. Experiments show that the proposed randomization techniques are helpful for big data anonymization and efficient for big data storage and computation.
\end{abstract}

\maketitle

\section{Introduction}
Tensor decomposition, as a multi-dimensional generalization of matrix decomposition, is a multi-decades-old mathematical technique in multi-way data analysis since the 1960s, see~\cite{cichocki2015tensor} and references therein; tensor decompositions are widely applied in areas from signal processing such as blind source separation and multi-modal data fusion to machine learning such as model compression and learning latent variable models~\cite{sidiropoulos2017tensor,anandkumar2014tensor}. Tensor computing recently emerges as a promising solution for big data processing due to its ability to model wide variety of data such as graphical, tabular, discrete, and continuous data~\cite{lahat2015multimodal,Sorber2015,zhou2016}; algorithms to cater for different data quality / veracity or missing data~\cite{song2017} and provide real-time analytics for big data velocity such as streaming analytics~\cite{sun2006,sun2008}; and able to capture the complex correlation structure in data with large volume and generate valuable insights for many big data distributed applications~\cite{cichocki2016tensor,cichocki2017tensor}. Tensor network computing, on the other hand, is a well-established technique among the numerical community; the technique provides unprecedented large-scale scientific computing with performance comparable to competing techniques such as sparse-grid methods~\cite{khoromskij2018tensor,khoromskij2012tensors}. Tensor network (TN) represents a data or tensor block in a sparsely-interconnected, low-order core tensors (typically $3^{rd}$-order or $4^{th}$-order tensors) and the functions by distributed, multilinear tensor operations. TN was first discovered in quantum physics in the 1990s to capture and model the multi-scale interactions 
between the entangled quantum particles in a parsimonious manner~\cite{orus2019tensor}. TN was then independently re-discovered in the 2000s by the numerical community and has found wide applications ranging from scientific computing to electronic design automation~\cite{khoromskij2018tensor,grasedyck2013literature,khoromskij2012tensors}.


Big data generated from sensor networks or Internet-of-Things are essential for machine learning, in particular deep learning, in order to train cutting-edge intelligent systems for real-time decision making and precision analytics. However, big data may contain proprietary information or personal information such as location, health, emotion, and preference information of individuals which requires proper encryption and access control to protect users' privacy. Symmetric and asymmetric key cryptosystems work by adding entropy / disorderliness into data using encryption algorithms and (pseudo-)random number generator so that unauthorized users cannot find pattern from the ciphertext and decipher them, however, higher computational cost is usually incurred with added functionality such as secure operations (addition / multiplication) in homomorphic encryption and asymmetric key distribution in public key encryption. The pain point of encryption nowadays is complicated key management and distribution especially when organizations or enterprises are undergoing digital transformation to complex computing environments such as multi- / hybrid-cloud and mobile environments. The field of secure multi-party computation (SMPC) originates from Yao\textquotesingle s garbled circuit in the 1980s where untrusted parties jointly compute a function without disclosing their private inputs~\cite{yao1986generate}. SMPC has evolved and adopts distributed trust paradigm in recent years given the complex computing environments, increasing attack surfaces, and recurring security breaches; the secret shares are now distributed among multiple computing nodes in order to be information-theoretically secure, i.e., secure against adversary with unbounded computational resources. SMPC computing primitives include secret sharing, garbled circuit, and homomorphic encryption, the supported secure operations are arithmetic, boolean, comparison, and bitwise operations; other secure building blocks that are routinely being used in SMPC are oblivious transfer, commitment scheme, and zero-knowledge proof~\cite{cramer2012secure,evans2018pragmatic}. It is well-known that fully homomorphic encryption~\cite{gentry2009fully} suffers from high computational complexity, making it not practical to compute complex functions during operational deployment; secret sharing and garbled circuit are expensive in terms of communication complexity and therefore routinely operate with low-latency networks, furthermore, garbled circuit involves symmetric encryption during the online phase. The communication complexity of existing SMPC protocols can incur runtime delay from an order of magnitude using local-area network (LAN) setting to several orders using wide-area network (WAN) setting. 

The quest for scalability calls for innovative data security solutions which not only simplify privacy management, but also provide seamless integration between privacy-preserving big data storage / communication and computation / sharing. We believe this requires introducing a new secure computation primitive that is based on distributed / dispersed tensor network computation. However, TN increases the functionality and performance of multi-party computation at the expense of security. In contrast to classical encryption and SMPC techniques which are based on modular arithmetic and works on fixed-point representations; TN naturally supports both floating-point and fixed-point arithmetics / operations. Furthermore, TN representations allow further compression unlike encrypted computation techniques, which generally increase the storage and communication overhead. Therefore, this generally makes encrypted computation not scalable for big data processsing; whereas data compression prior to encryption usually makes the data representations lose some functionalities such as encrypted computation on the original data. With the impressive track records of distributed TNs in large-scale scientific computing and big data analytics, we propose a novel secret-sharing scheme based on tensor networks and investigate its feasibility for privacy-preserving big data distributed applications. Our contributions are as follows:
\begin{itemize}
\item Propose an arithmetic secret-sharing scheme based on randomized tensor network decomposition and dispersed tensor multilinear operations. The randomization is done by controlled perturbation applied to the data blocks prior to singular value decomposition (SVD), which results in randomized tensor blocks after decomposition due to the complex structural information in big data. The perturbation technique can be easily adapted in various TN decomposition algorithms to generate randomized TN representations.
\item Empirically analyze the privacy leakage of the randomized TN representations for 1D to 3D datasets and propose mitigation techniques to reduce the privacy leakage. The data compressibility and algorithmic efficiency of the proposed randomized TN algorithms have also been investigated.
\end{itemize}
The organization of this work is as follows: Section~\ref{sec:rel_work} covers related work on state-of-the-art privacy-preserving techniques and secure tensor decompositions. Sections~\ref{sec:threat_model_sec} and~\ref{sec:tn_data_privacy} explain the security model and our proposed randomized tensor dispersed computing approach for big data privacy preservation. Section~\ref{sec:exp} conducts experimental studies to benchmark the security, efficiency, and performance of the proposed approach. Section~\ref{sec:disc} discusses the implications, limitations, and potential extension of this research study.  

\section{Related Work}
\label{sec:rel_work}

\emph{Secret-Sharing} schemes provide information-theoretical security at the expense of high storage and communication cost. Here, we review practical secret-sharing schemes for big data protection that provide only computational security but offer high storage / computational efficiency. Krawczyk~\cite{krawczyk1993secret} proposes the first computational secret sharing scheme by encrypting the data using symmetric encryption with randomly-generated key, the encrypted data is divided into multiple blocks using Rabin's information dispersal algorithm; whereas the encryption / decryption key is split using Shamir's secret-sharing scheme such that collecting a certain threshold number of blocks is enough for secret reconstruction. Since then, many variants of the computational secret-sharing scheme have been proposed to improve the data security, data redundancy / error resistance, performance, data integrity / authentication, fragment size, data deduplication, and location management with different machine trustworthiness~\cite{kapusta2015data,memmi2015data,kapusta2017data}. Most notably, the key exposure problem is a practical issue to address due to usage of weak key for encryption, key reuse, or key leakage. The All-Or-Nothing Transform (AONT) introduced by Rivest~\cite{rivest1997all} solves the key exposure problem by building dependency between the fragments such that acquiring only the key without all the fragments will not lead to immediate information leakage, a recent review on AONT can be found in~\cite{qiu2019all}. Furthermore, access revocation is greatly simplified by re-encrypting only one data fragment with a fresh encryption key, which significantly reduces the transmission cost~\cite{kapusta2019poster,kapusta2019secure}. However, utility of such encrypted data is quite limited such as search, update, and computation cannot be performed without reconstructing the original data~\cite{fabian2015collaborative,yuksel2017research}.

\emph{Database Fragmentation or Data Splitting}~\cite{domingo2019privacy} aim to provide functionality-preserving data protection for data storage on clouds. Sensitive data is fragmented in clear form in separate storage locations such that each data fragment does not reveal confidential information linked to a subject. Data splitting can be done at byte, semantic, or attribute level. Byte-level fragmentation splits the sensitive files and performs shifting and recombination of the bytes to form fixed data blocks before storing on different cloud locations, this is particularly suitable for binary or multimedia files, which are usually stored but not processed by cloud. Semantically-grounded splitting mechanism is well-suited for unstructured data such as textual data, it can provide keyword search for online document, email, and messaging applications. For example, a recent work by~\cite{sanchez2017privacy} automatically detects and splits the sets of textual entities that may disclose sensitive information by analysing the semantics they convey and their semantic dependencies. Attribute-level splitting such as vertical splitting~\cite{aggarwal2005two} is very useful for statistical databases because usually is the combination of several risky attributes that may lead to personal re-identification. Computation on attributes stored on single fragment in vertial splitting is fast and straightforward, e.g., addition, updating, and uni-valued statistics such as mean and variance. However, data splitting requires a proxy server to manage the locations, queries, and operations on the data fragments, this becomes the single point of failure if users cannot access the metadata stored at the proxy.

\emph{Data Anonymization} is perhaps the simplest low-cost solution that is widely adopted nowadays for secure data sharing within and across enterprises for diverse applications, including machine learning. Data anonymization techniques cover both the removal of personally-idenfiable information (e.g., using hashing or masking techniques) and data randomization / perturbation techniques (e.g., random noise, permutation, transformation)~\cite{domingo2019privacy}. The random components or functions have to be carefully designed to preserve important information in the training dataset and ensure model performance. A recent systematic survey of different privacy metrics that have been proposed over the years can be found in ~\cite{wagner2018technical}. These privacy metrics are based on information theory, data similarity, indistinguishability measures, and adversary's success probability; to choose a suitable privacy metric for a particular setting depends on the adversarial model, data sources, information available to compute the metric and the properties to measure~\cite{wagner2018technical}. Differential privacy (DP)~\cite{dwork2014algorithmic, dwork2011differential} is a mathematical framework to rigorously quantify the amount of information leaked during operations on a statistical database or machine learning~\cite{sarwate2013signal,chaudhuri2011differentially,bassily2014private,abadi2016deep,shokri2015privacy}, DP is a proven privacy-preserving technique widely adopted by the industry. A recent promising data anonymization approach is to generate synthetic data~\cite{nikolenko2019synthetic} that resembles the statistical distribution or behavior observed in the original datasets using generative machine learning models such as generative adversarial networks~\cite{goodfellow2014generative} and computer simulations (e.g.,~\cite{lopez2016review}), however, these models / simulations are application-specific (i.e., depend on the training dataset or physical models) and any analysis on the synthetic data has to be verified over the real dataset for validation.

Although privacy-preserving matrix and tensor decomposition techniques have been well studied in the literature~\cite{nikolaenko2013privacy,liu2015fast,berlioz2015applying,friedman2016differential,zhang2018jo,wang2016online,imtiaz2018distributed,ma2019privacy,kim2017federated,feng2018privacy,kuang2015secure}, distributed / dispersed TN representations and computation have not been proposed for big data privacy preservation, which motivates the current study. Different from data anonymization techniques, tensor decompositions are fully reversible and compressible, the reconstruction accuracy can be either lossy or near-lossless~\cite{dauwels2012near}. Unlike data splitting, TN does not require proxy server to manage the metadata, but offers much better utility of the decomposed data at the expense of higher privacy leakage compared to computational secret-sharing schemes. To process big data, randomized mapping or projection techniques utilize a projection matrix such as Gaussian, Rademacher, and random orthonormal matrices~\cite{caiafa2014stable,tropp2017randomized} to project the data tensor to much smaller tensor size before applying tensor decompositions. Randomized sampling techniques such as fiber subset selection or tensor cross approximation choose a small subset of tensor fibers that approximate the entire data tensor well, e.g., measured using quasi-optimal maximal volume or modulus determinant of the submatrix so that the matrix cross-approximation is close to the optimal SVD solution~\cite{mahoney2009cur,oseledets2010tt}. Existing randomized mapping / projection and randomized sampling algorithms are useful for big data tensor decompositions to fit the data size into existing memory requirements, the decomposed tensor blocks are usually compressed with lossy reconstruction accuracy, which is different from our proposed randomized tensor decompositions. The randomness of the decomposed tensor blocks is also limited by the distribution of the projection matrix and sampling process to ensure small error bounds, whereas our proposed algorithms randomly disperse the complex structural information of big data into the tensor cores by applying large-but-controlled perturbations during the sequential matrix decomposition process in tensor decomposition algorithms. The time complexity is also much lower compared to randomized projection / mapping algorithms and can be easily adapted into existing TN algorithms. Nonetheless, the proposed tensor perturbation techniques can be easily combined with existing randomized projection / sampling algorithms for big data processing and privacy protection. Tensor decompositions have been widely used for dimensionality reduction of big data, however, research on tensor network coding schemes are lagging behind, only a few publications are found at the time of writing~\cite{dauwels2012near,karami2012compression,ballester2019tthresh}.

\section{Threat Model and Security}
\label{sec:threat_model_sec}

The secure storage and computation by a client are outsourced to a set of untrusted but non-colluding servers $S_1, S_2, ..., S_n$, the client secret share their inputs among the servers in the initial setup phase, the servers then proceed to securely store (e.g., with encryption), compute and communicate using dispersed TN computation protocols. The servers run on different software stacks to minimize the chance that they all become vulnerable to the exploit available to malware attacks and can be operated under different sub-organizations to minimize insider threats. Given the cloud scenario, the secret shares can be distributed to multiple virtual instances provided by the same cloud service provider (CSP) or to different clouds (e.g., multi-cloud or hybrid-cloud environments). We assume a semi-honest adversary $\mathbb{A}$ (or so-called honest-but-curious adversary) who is able to corrupt any subset of the clients and at most $n-1$ servers at any point of time. Different from encrypted data processing, our security definition requires an adversary to learn only partial information of the client's input but not knowing the sensitive information from the process. The privacy leakage is measured based on information-theoretic and similarity-based privacy metrics. Secret-sharing scheme based on TN is asymmetric to each server, i.e., each server contains index-specific information. As shown in Sections~\ref{sec:tn_data_privacy} and~\ref{sec:exp}, each of the TN representations requires high data complexity (or high tensor-rank complexity) to be privacy-preserving in multi-party setting.

\section{Secret-Sharing Scheme Based on Distributed Tensor Networks}
\label{sec:tn_data_privacy}

In this section, we propose a novel secret-sharing scheme based on dispersed TN representations / operations to seamlessly secure big data storage, communication, sharing, and computation. TN decomposes data chunk at the semantic level, each of the decomposed tensor block which contains latent information are randomly distributed among multiple non-colluding servers. The success of multi-way component analysis can be attributed to the existence of efficient algorithms for matrix and tensor decomposition and the possibility to extract components with physical meaning by imposing constraints such as sparsity, orthogonality, smoothness, and non-negativity~\cite{cichocki2016tensor}. Our primary intuition is that higher-order tensor decompositions are in general non-unique, each tensor core or factor matrix contains index-specific information which are unlinkable and uninterpretable due to non-uniqueness of the decompositions, therefore they are commonly used for dimensionality reduction and compressed computation.

Several basic tensor models are described here within the multi-party computation setting to enhance the privacy protection of the original tensor. Here, we propose randomized algorithm based on perturbation technique to decompose each data chunk into randomized tensor blocks, each of the tensor blocks can be re-randomized again using tensor-rounding algorithm after performing tensor distributed, multilinear operations to reduce the tensor-rank complexity for storage and computational efficiency.

\emph{Tucker decomposition (TD)}~\cite{papalexakis2016tensors} is a natural extension of matrix Singular Value Decomposition (SVD) into high-dimensional tensor. TD captures the interactions between the latent factors $\textbf{U}$ (from SVD of mode-n matricization of a tensor) using a core tensor $\mathcal{G}$ that reflects and ranks the major subspace variations in each mode of the original tensor. For a third-order tensor $\mathcal{A}\in\mathbb{R}^{I_1\times I_2\times I_3}$, TD can be defined as follows using different tensor operations:
\begin{equation}
\begin{split}
\mathcal{A}(i_1,i_2,i_3)&\cong \mathcal{G}\times_1 \langle \textbf{U}_1\rangle_1 \times_2 \langle \textbf{U}_2\rangle_2 \times_3 \langle \textbf{U}_3\rangle_3\\
vec(\mathcal{A})&\cong (\langle \textbf{U}_3\rangle_3\otimes \langle \textbf{U}_2\rangle_2\otimes \langle \textbf{U}_1\rangle_1)\hspace{0.1cm}vec(\mathcal{G})
\end{split}
\end{equation}
$\mathcal{G}\in\mathbb{R}^{R_1\times R_2\times R_3}$ is a 3-dimensional core tensor, $\textbf{U}_k\in\mathbb{R}^{I_k\times R_k}$, $k\in\{1,2,3\}$ are the factor matrices, $\times_n$ is the n-mode product, $\otimes$ is the Kronecker product, $vec(\cdot)$ is the vectorization operator (see the definitions in~\cite{papalexakis2016tensors}), $\langle\cdot\rangle_\ell$ denotes the private share stored in server $\ell$. Here, $\mathcal{G}$ is a shared core for exchange between servers to perform tensor computation schemes. TD is non-unique because the latent factors can be rotated without affecting the reconstruction error, however, TD yields a good low-rank approximation of a tensor in terms of squared error. Canonical Polyadic (CP) decomposition is a special case of TD when $\mathcal{G}$ is superdiagonal. CP is very popular in signal processing due to its uniqueness guerantee and ease of interpretation~\cite{cichocki2015tensor}, however, these properties also make CP unsuitable for privacy preservation.

\emph{Hierarchical Tucker (HT) decomposition}~\cite{hackbusch2009new,grasedyck2010hierarchical} was proposed to reduce the memory requirements of TD. HT approximates well higher-order tensors ($N>>3$) without suffering from the curse of dimensionality. HT recursively splits the modes of a tensor based on a binary tree hierarchy such that each node contains a subset of the modes. Therefore, HT requires a priori knowledge of a binary tree of matricizations of the tensor, HT is defined as follows:
\begin{equation}
\textbf{U}_t\cong ( \textbf{U}_{t_l}\otimes \textbf{U}_{t_r})\hspace{0.05cm}\langle \textbf{B}_t\rangle_t
\end{equation}
$\textbf{B}_t$ are the ``transfer" core tensors (or internal nodes) reshaped into $R_{t_l}R_{t_r}\times R_t$ matrix, $\textbf{U}_t$ contains the $R_t$ left singular vectors of the original tensor, $t_l$ and $t_r$ correspond to the left and right child nodes respectively. The leaf nodes $\langle \textbf{U}_1\rangle_1,\langle \textbf{U}_2\rangle_2,\ldots,\langle \textbf{U}_N\rangle_N$ contain the latent factors and should be stored distributedly to ensure privacy preservation. HT is particularly useful when the application provides an intuitive and natural hierarchy over the physical modes.

\emph{Tensor-Train (TT)}~\cite{oseledets2011tensor} decomposes a given tensor into a series or cascade of connected core tensors, therefore TT can be interpreted as a special case of HT. TT core tensors are connected through a common reduced mode or TT-rank, $R_k$. TT is defined as follows:
\begin{equation}
    \mathcal{A}(i_1,i_2,i_3)\cong\langle \textbf{G}[i_1]\rangle_1\times \langle \textbf{G}[i_2]\rangle_{2} \times\langle \textbf{G}[i_3]\rangle_3
\end{equation}
where $\textbf{G}[i_k]$ is a $R_{k-1}\times R_k$ matrix with $R_0=R_3=1$, and $\times$ is the matrix multiplication operation. TT format and its variants are very useful owing to their flexibilty for a number of distributed, multilinear operations~\cite{lee2018fundamental} and the possibility to convert other basic tensor models (e.g., CP, TD, HT) into TT format~\cite{cichocki2016tensor}. Similar properties apply to tensor chain or tensor ring format (TR)~\cite{zhao2016tensor}, which is a linear combination of TT formats, i.e., $R_1=R_3>1$. TR representations are more generalized and powerful compared to TT representations~\cite{zhao2016tensor}; whereas extended TT further decomposes the TT-cores into smaller blocks~\cite{hackbusch2012tensor}. 

\emph{Storage Complexity.} Table~\ref{tab:storage_comp} tabulates the storage complexity and bound of the different TN formats mentioned here. Low-rank approximation is very useful in tensor network computing for saving storage, communication, and computational cost with negligible loss in accuracy for some highly-correlated tensor data structures or functional forms that admit low-rank structure. Tucker format is not practical for tensor order $N>5$ because the number of entries of the core tensor $\mathcal{G}$ scales exponentially with $N$, therefore storage and computing in Tucker format are not practical when dealing with higher-order tensors~\cite{cichocki2016tensor}. TT format and its variants exhibit both stable numerical properties and reasonable storage complexity. Furthermore, TT allows control of the approximation error within the TT decomposition and TT-rounding algorithms.

\begin{table}[ht]
\begin{center}
\caption{Storage complexity of different tensor formats~\cite{cichocki2016tensor}. The storage bound is calculated by letting $I=\max_k I_k$, $R=\max_k R_k$, $k\in\{1,2,\ldots,N\}$. HT, TT, and TR are powerful representations that break the curse of big data dimensionality.}
\label{tab:storage_comp}
\begin{tabular}{ |c|c|c| } 
 \hline
 \textbf{TN}&\textbf{Storage Complexity}&\textbf{Storage Bound} \\\hline 
 CP &$\sum_{k=1}^N I_k R$&$O(NIR)$ \\\hline 
 TD &$\sum_{k=1}^N I_k R_k+\prod_{k=1}^N R_k$&$O(NIR+R^N)$ \\\hline
 HT &$\sum_{k=1}^N I_k R_k + \sum_{(u,v,t)} R_u R_v R_t$&$O(NIR+NR^3)$ \\\hline
 TT / TR &$\sum_{k=1}^{N}I_k R_{k-1} R_{k}$&$O(NIR^2)$ \\\hline
\end{tabular}
\end{center}
\end{table}

\begin{figure}
\centering
\includegraphics[width=0.95\linewidth]{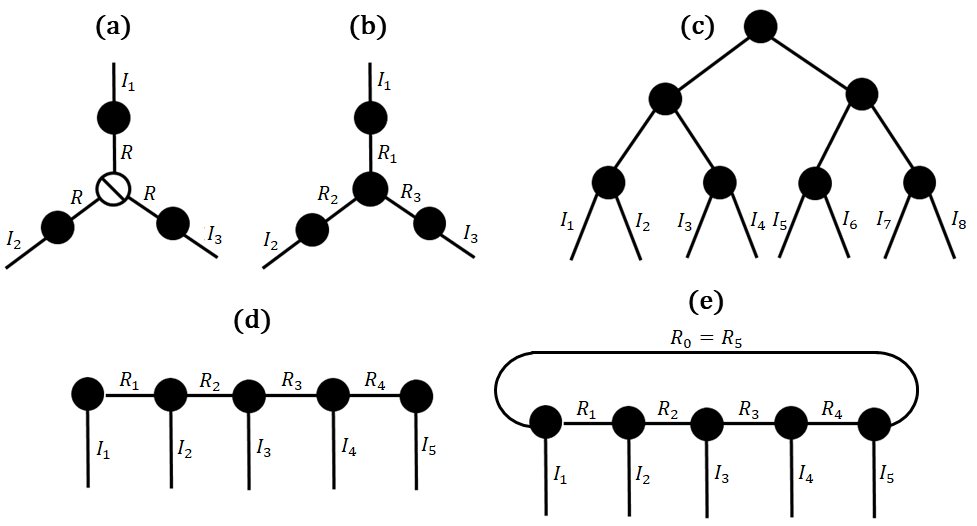}
\caption{Graphical representations of different tensor network (TN) decompositions. The number of lines connected to a node shows the tensor order; the rank and mode size are labeled on the edges. (a) Canonical Polyadic (CP) decomposition, (b) Tucker decomposition (TD), (c) Hierarchical Tucker (HT), (d) Tensor-Train (TT), and (e) Tensor-Ring (TR) decomposition. A TN can be partitioned into secret shares at individual node level or into sets of tensor nodes distributed across servers for privacy preservation.}
\label{fig:tensor_network}
\end{figure}

\emph{Graphical Representations.} TNs can be represented by a set of nodes interconnected by the edges. The edges correspond to the contracted modes, whereas lines that do not go from one tensor to another correspond to open (physical) modes, which contribute to the total number of orders of the entire TN. Fig.~\ref{fig:tensor_network} shows the graphical representations of different TN representations. Mathematical operations performed on tensor (e.g., tensor contractions and reshaping) can be expressed using graphical representation of tensors in a simple and intuitive way  without the explicit use of complex mathematical expressions.

\begin{figure}[ht]
  \centering
  \includegraphics[width=\linewidth]{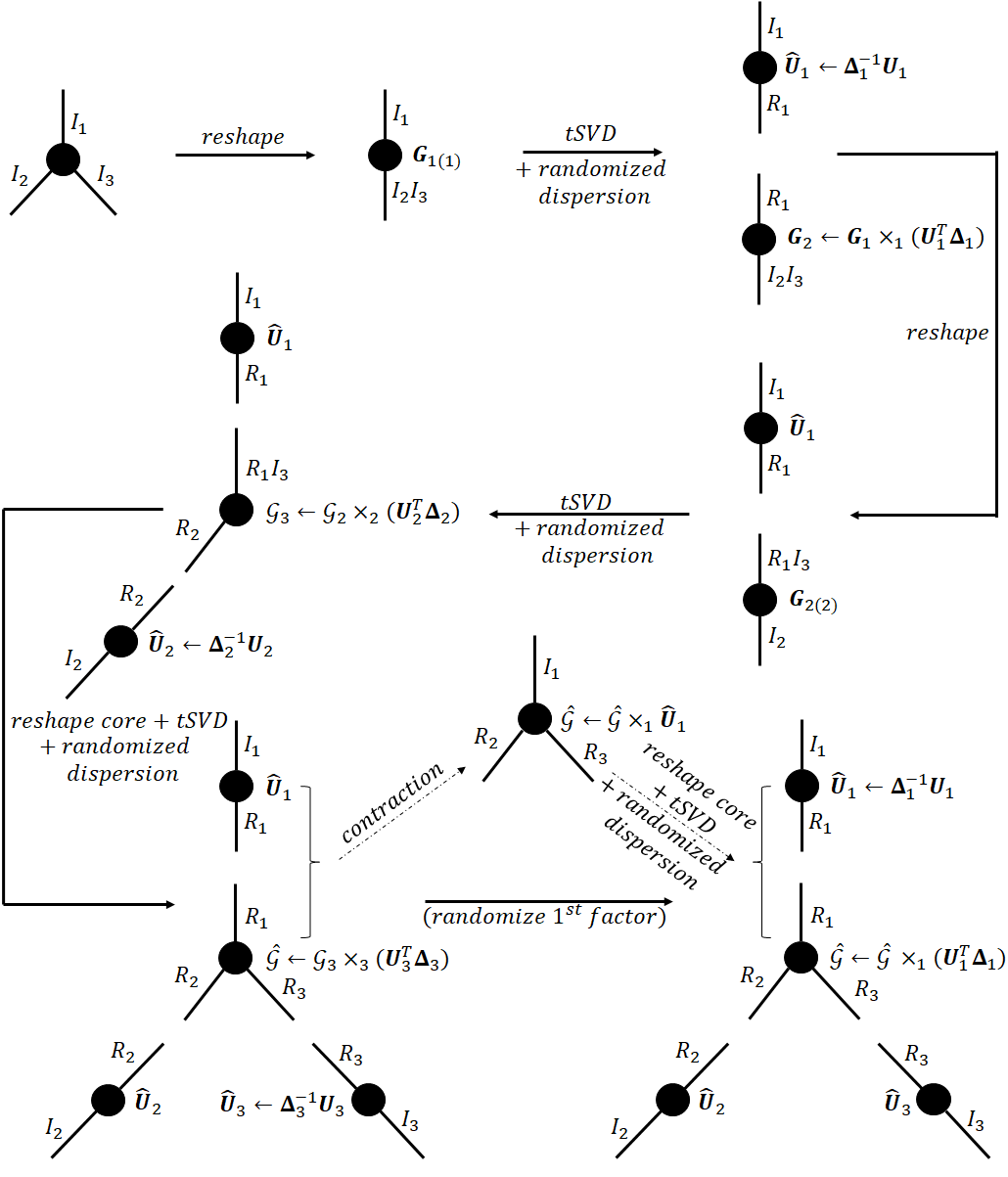}
  \caption{Graphical representation of the proposed rTD algorithm for a $3^{rd}$-order tensor, see Algorithm~\ref{alg:rand_td_svd} for the details.}
  \label{fig:rand_td_svd}
\end{figure}

\begin{figure}[ht]
  \centering
  \includegraphics[width=0.95\linewidth]{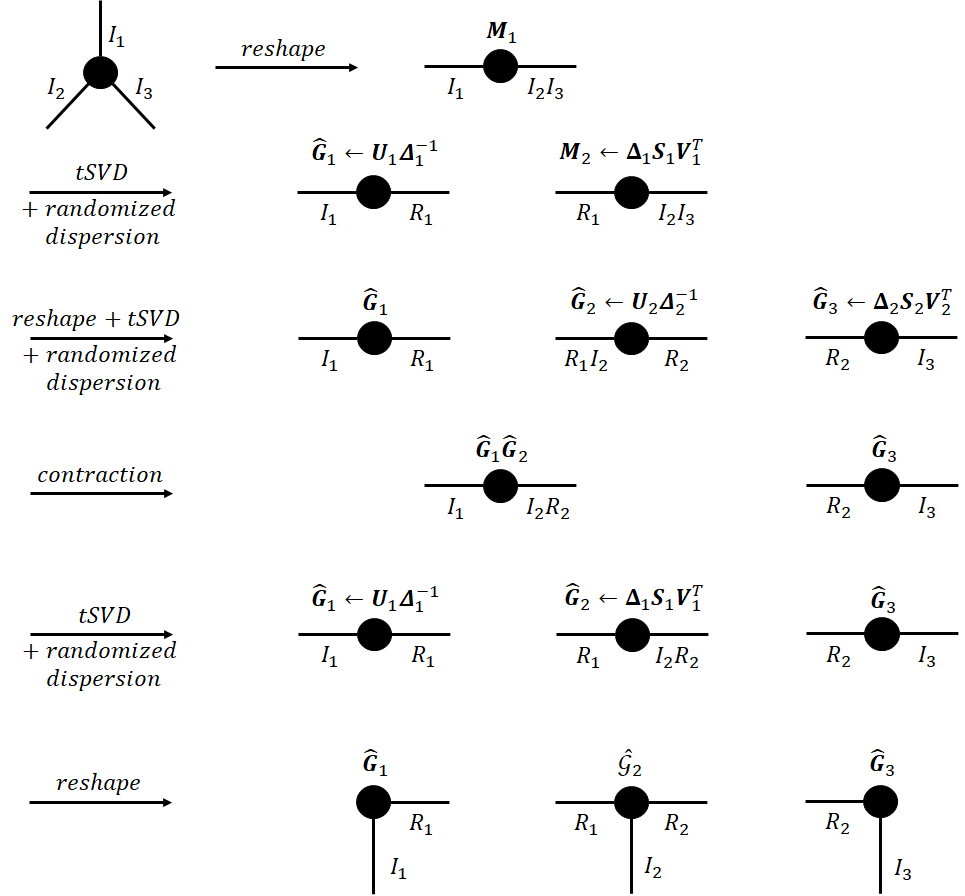}
  \caption{Graphical representation of the proposed rTT-SVD for a $3^{rd}$-order tensor, see Algorithm~\ref{alg:rand_tt_svd} for the details.}
  \label{fig:rand_tt_svd}
\end{figure}

\emph{Shares Generation based on Randomized Tensor Decompositions.} Algorithms~\ref{alg:rand_td_svd},~\ref{alg:rand_ht_svd},~\ref{alg:rand_tt_svd}, and~\ref{alg:rand_tr_svd} present our proposed randomized rTD, rHT, rTT-SVD, rTR-SVD algorithms that decompose N-dimensional tensor into randomized secret shares by applying perturbations to randomly disperse the structural information in big data into the tensor cores. Algorithm~\ref{alg:rand_td_svd} is based on Higher-Order Singular Value Decomposition (HOSVD) proposed in~\cite{bergqvist2010higher}, HOSVD performs SVD on each mode of a tensor to extract the latent factors before obtaining the core tensor that captures the complex interactions between the latent factors. Algorithm~\ref{alg:rand_ht_svd} recursively applies rTD on each tensor node based on a binary tree matricizations of the input tensor. Algorithm~\ref{alg:rand_tt_svd} and~\ref{alg:rand_tr_svd} are based on the TT-SVD and TR-SVD algorithms proposed in~\cite{oseledets2011tensor} and~\cite{zhao2016tensor} respectively, TT-SVD and TR-SVD perform sequential SVD decomposition on a tensor to obtain the TT and TR representations. Figures~\ref{fig:rand_td_svd} and~\ref{fig:rand_tt_svd} show the graphical representations of the proposed rTD and rTT-SVD algorithms. The randomized dispersion is applied after performing each SVD step in Algorithms~\ref{alg:rand_td_svd},~\ref{alg:rand_ht_svd}, and~\ref{alg:rand_tt_svd}. To balance between compression and randomness, the maximum (randomized) perturbation should be within certain threshold $\delta$ based on the magnitude of each singular value, and +ve/-ve sign difference from each singular vector. The share re-generation can be done with our proposed randomized TT-rounding algorithm based on~\cite{oseledets2011tensor} (see Algorithm~\ref{alg:rand_tt_round}) all carried out in TT format on distributed servers, however this is not recommended because computation with TN may leak private information (gradually) to the servers. The proposed secret-sharing scheme is asymmetric to the servers, each server stores only index-specific information based on the tensor core it receives. The perturbations are embedded inside existing tensor decomposition algorithms, therefore the computational complexity does not increase much, only a few more tensor core contractions (i.e., to apply perturbation and randomize $1^{st}$ core / factor) and an SVD are performed. The memory size to store the perturbation factors is considered negligible.

\begin{algorithm}
    \SetAlgoLined
    \SetKwInOut{Input}{Input}
    \SetKwInOut{Output}{Output}
    \SetKwInOut{Return}{Return}
    \Input{Tensor $\mathcal{A}\in\mathbb{R}^{I_1\times I_2\times\ldots\times I_N}$ and ranks $R_1,R_2,\ldots,R_N$.}
    \Output{Tucker core $\hat{\mathcal{G}}\in\mathbb{R}^{R_1,R_2,\ldots,R_N}$ and factor matrices $\hat{\textbf{U}}_k\in\mathbb{R}^{I_k\times R_k}\hspace{0.1cm}s.t.\hspace{0.1cm} \mathcal{A}\cong\hat{\mathcal{G}}\times_1\hat{\textbf{U}}_1\times_2\hat{\textbf{U}}_2\ldots\times_N\hat{\textbf{U}}_N$.}
    Initialization:$\hspace{0.1cm}\mathcal{G}_1=\mathcal{A}$;\\
    \textbf{Modified from multilinear SVD or $N$-mode SVD:}\\
    \For{$k=1$ to $N$}{
        $\left[\textbf{U}_k,\textbf{S}_k,\textbf{V}_k\right]\leftarrow tSVD(\textbf{G}_{k(k)},R_{trunc.}=R_k)$;\\
        Generate diagonal perturbation matrix with uniform distribution bet. threshold $\delta$ and $1$, $\pmb{\Delta}_k\sim\mathcal{U}([\delta,1])$;\\
        Perturb the core tensor, $\mathcal{G}_{k+1}\leftarrow \mathcal{G}_k\times_k (\textbf{U}^T_k\pmb{\Delta}_{k})$;\\
        Update the factor matrix, $\hat{\textbf{U}}_k\leftarrow \pmb{\Delta}^{-1}_k\textbf{U}_k$;
    }
    \textbf{Randomize the $1^{st}$ TD factor matrix:}\\
    $\quad\hat{\mathcal{G}}\leftarrow \mathcal{G}_{N+1}$;$\hspace{0.1cm}\hat{\mathcal{G}}\leftarrow\hat{\mathcal{G}}\times_1\hat{\textbf{U}}_1$;\\
    $\quad[\textbf{U}_1,\textbf{S}_1,\textbf{V}_1]\leftarrow tSVD(\hat{\textbf{G}}_{(1)},R_{trunc.}=R_1)$;\\
    $\quad\pmb{\Delta}_1\sim\mathcal{U}([\delta,1]);\hat{\mathcal{G}}\leftarrow\hat{\mathcal{G}}\times_1 (\textbf{U}^T_1\pmb{\Delta}_1)$;\\
    $\quad\hat{\textbf{U}}_1\leftarrow \pmb{\Delta}^{-1}_1\textbf{U}_1$;\\
    \caption{Proposed randomized Tucker Decomposition (rTD) based on Higher-Order SVD (HOSVD)~\cite{bergqvist2010higher}.}
    \label{alg:rand_td_svd}
\end{algorithm}

\begin{algorithm}
    \SetAlgoLined
    \SetKwInOut{Input}{Input}
    \SetKwInOut{Output}{Output}
    \SetKwInOut{Return}{Return}
    \Input{Tensor $\mathcal{A}\in\mathbb{R}^{I_1\times I_2\times\ldots\times I_N}$, ranks $R_1,R_2,\ldots,R_N$, and binary tree $\mathcal{T}$ of the matricizations of $\mathcal{A}$.}
    \Output{HT factor matrices $\hat{\textbf{U}}_1,\hat{\textbf{U}}_2,\ldots,\hat{\textbf{U}}_N$ and transfer cores $\hat{\mathcal{B}}_t$, $t\in$ nonleaf nodes of binary tree $\mathcal{T}$.}
    $\textbf{U}_{1}\leftarrow \textbf{A}_{(1)}$;\\
    \textbf{Starting from the root node of tree} $\mathcal{T}$\textbf{, select a node} $t$:\\
    $\quad$ Set $t_l$ and $t_r$ to be the left and right child of $t$ resp.;\\
    $\quad$ If $t_l$ is not singleton: $R_{t_l}\leftarrow R_{full}$;\\
    $\quad$ If $t_r$ is not singleton: $R_{t_r}\leftarrow R_{full}$;\\
    $\quad\mathcal{U}_t\leftarrow reshape(\textbf{U}_t,[t_l,t_r,t])$;\\
    $\quad[\hat{\mathcal{B}}_t, \textbf{U}_{t_l}, \textbf{U}_{t_r}]\leftarrow rTD(\mathcal{U}_{t}, R_{trunc.}=[R_{t_l},R_{t_r}])$;\\
    $\quad$ If $t_l$ is a singleton: $\hat{\textbf{U}}_{t_l}\leftarrow \textbf{U}_{t_l}$;\\
    $\quad$ If $t_r$ is a singleton: $\hat{\textbf{U}}_{t_r}\leftarrow \textbf{U}_{t_r}$;\\
    $\quad$ Recurse on $t_l$ and $t_r$ until $t_l$ and $t_r$ are singletons.
    \caption{Proposed randomized Hierarchical Tucker (rHT) decomposition by recursive node-wise rTD (Algo.~\ref{alg:rand_td_svd}).}
    \label{alg:rand_ht_svd}
\end{algorithm}

\begin{algorithm}
    \SetAlgoLined
    \SetKwInOut{Input}{Input}
    \SetKwInOut{Output}{Output}
    \SetKwInOut{Return}{Return}
    \Input{Tensor $\mathcal{A}\in\mathbb{R}^{I_1\times I_2\times\ldots\times I_N}$ and error threshold $\epsilon$.}
    \Output{TT cores $\hat{\mathcal{A}}=\hat{\textbf{G}}_1\cdot\hat{\mathcal{G}}_2\cdots\hat{\mathcal{G}}_{N-1}\cdot\hat{\textbf{G}}_N$ such that the approximation error $\lVert \mathcal{A}-\hat{\mathcal{A}} \rVert_F\lesssim\epsilon\lVert \mathcal{A}\rVert_F$.}
    Initialization: TT-rank $R_0=1$; Perturbation threshold $\delta$; \\\hspace{1.8cm}Truncation parameter $\delta_\epsilon = \frac{\epsilon}{\sqrt{N-1}}$;\\
    Tensor shape, $[I_1,I_2,\ldots,I_N]\leftarrow shape(\mathcal{A})$;\\
    Mode-1 matricization of tensor $\mathcal{A}$, $\textbf{M}_1\leftarrow \textbf{A}_{(1)}$;\\
    \textbf{Sequential (SVD $+$ randomized dispersion):}\\
    \For{$k=1$ to $N-1$}{
        Truncated SVD, $\left[\textbf{U}_k,\textbf{S}_k,\textbf{V}_k\right]\leftarrow tSVD(\textbf{M}_k,\delta_{rel.}=\delta_\epsilon)$;\\
        TT-rank, $R_k\leftarrow shape(\textbf{S}_k, 1)$;\\
        Generate diagonal perturbation matrix with uniform dist. between threshold $\delta$ and $1$, $\pmb{\Delta}_k\sim\mathcal{U}([\delta,1])$;\\
        Reshape the orthogonal matrix $\textbf{U}_k$ divided by the perturbation factor $\pmb{\Delta}_k$ into a third-order tensor
        $\hat{\mathcal{G}}_k\leftarrow reshape(\textbf{U}_k\pmb{\Delta}^{-1}_k,[R_{k-1},I_k,R_k])$;\\
        Matricize $\textbf{S}_k\textbf{V}_k^T$ multiplied by the perturbation factor $\pmb{\Delta}_k$
        $\textbf{M}_{k+1}\leftarrow reshape(\pmb{\Delta}_k\textbf{S}_k\textbf{V}^T_k,[R_kI_{k+1},\prod_{p=k+2}^{N}I_p])$;\\
    }
    Construct the last core, $\hat{\textbf{G}}_N\leftarrow reshape(\textbf{M}_N,[R_{N-1},I_N])$\;
    \textbf{Randomize the $1^{st}$ TT-core:}\\
    $\quad \hat{\textbf{G}}_1\leftarrow reshape(\hat{\mathcal{G}}_1,[I_1,R_1])$;\\
    $\quad \hat{\textbf{G}}_2\leftarrow reshape(\hat{\mathcal{G}}_2,[R_1,I_2R_2])$;\\
    $\quad [\textbf{U}_1,\textbf{S}_1,\textbf{V}_1]\leftarrow tSVD(\hat{\textbf{G}}_1\hat{\textbf{G}}_2,\delta_{rel.}=\delta_\epsilon)$;\\
    $\quad R_1\leftarrow shape(\textbf{S}_1, 1)$; $\pmb{\Delta}_1\sim\mathcal{U}([\delta,1])$;\\
    $\quad\hat{\textbf{G}}_1\leftarrow reshape(\textbf{U}_1\pmb{\Delta}^{-1}_1,[I_1,R_1])$;\\
    $\quad\hat{\mathcal{G}}_2\leftarrow reshape(\pmb{\Delta}_1\textbf{S}_1\textbf{V}_1^T,[R_1,I_2,R_2])$;\\
    \caption{Proposed randomized Tensor Train-Singular Value Decomposition (rTT-SVD) algorithm based on~\cite{oseledets2011tensor}.}
    \label{alg:rand_tt_svd}
\end{algorithm}

\begin{algorithm}
    \SetAlgoLined
    \SetKwInOut{Input}{Input}
    \SetKwInOut{Output}{Output}
    \SetKwInOut{Return}{Return}
    \Input{Tensor $\mathcal{A}\in\mathbb{R}^{I_1\times I_2\times\ldots\times I_N}$ and error threshold $\epsilon$.}
    \Output{TR cores $\hat{\mathcal{A}}=\hat{\mathcal{G}}_1\cdot\hat{\mathcal{G}}_2\cdots\hat{\mathcal{G}}_N$ such that the approximation error $\lVert \mathcal{A}-\hat{\mathcal{A}} \rVert_F\lesssim\epsilon\lVert \mathcal{A}\rVert_F$.}
    Initialization: Perturbation threshold $\delta$; \\\hspace{1.8cm}Truncation parameter $\delta_k = \begin{cases} \frac{\sqrt{2}\epsilon}{\sqrt{N}},k=1\\
    \frac{\epsilon}{\sqrt{N}},k>1\\
    \end{cases}$;\\
    \textbf{Prepare the $1^{st}$ TR core:}\\
    $\quad$ Tensor shape, $[I_1,I_2,\ldots,I_N]\leftarrow shape(\mathcal{A})$;\\
    $\quad$ Mode-1 matricization of tensor $\mathcal{A}$, $\textbf{M}_1\leftarrow \mathcal{A}_{(1)}$;\\
    $\quad$ Truncated SVD, $\left[\textbf{U}_1,\textbf{S}_1,\textbf{V}_1\right]\leftarrow tSVD(\textbf{M}_1,\delta_{rel.}=\delta_1)$;\\
    $\quad$ $\pmb{\Delta}_1\sim\mathcal{U}([\delta,1])$; Split TT-ranks $R_0,R_1$:\\ $\quad\quad\min_{R_0,R_1} \vert\vert R_0-R_1 \vert\vert$ s.t. $R_0R_1=shape(\textbf{S}_1,1)$;\\
    $\quad$ Set TT-rank, $R_N\leftarrow R_0$;\\
    $\quad\hspace{0.1cm} \hat{\mathcal{G}}_1\leftarrow reshape(\textbf{U}_1\pmb{\Delta}^{-1}_1,[R_{0},I_1,R_1])$;\\
    $\quad\hspace{0.1cm} \textbf{M}_{2}\leftarrow reshape(\pmb{\Delta}_1\textbf{S}_1\textbf{V}^T_1,[R_1I_{2},\prod_{p=3}^{N}I_pR_N])$;\\
    \textbf{Sequential (SVD $+$ randomized dispersion):}\\
    \For{$k=2$ to $N-1$}{
        $\left[\textbf{U}_k,\textbf{S}_k,\textbf{V}_k\right]\leftarrow tSVD(\textbf{M}_k,\delta_{rel.}=\delta_k)$;\\
        $R_k\leftarrow shape(\textbf{S}_k, 1)$; $\pmb{\Delta}_k\sim\mathcal{U}([\delta,1])$;\\
        $\hat{\mathcal{G}}_k\leftarrow reshape(\textbf{U}_k\pmb{\Delta}^{-1}_k,[R_{k-1},I_k,R_k])$;\\
        $\textbf{M}_{k+1}\leftarrow reshape(\pmb{\Delta}_k\textbf{S}_k \textbf{V}^T_k,[R_kI_{k+1},\prod_{p=k+2}^{N}I_pR_N])$;\\
    }
    Construct the last core, $\hat{\mathcal{G}}_N\leftarrow reshape(\textbf{M}_N,[R_{N-1},I_N,R_N])$\;
    \textbf{Randomize the $1^{st}$ TR core:}\\
    $\quad \hat{\textbf{G}}_1\leftarrow reshape(\hat{\mathcal{G}}_1,[R_0I_1,R_1])$;\\
    $\quad \hat{\textbf{G}}_2\leftarrow reshape(\hat{\mathcal{G}}_2,[R_1,I_2R_2])$;\\
    $\quad [\textbf{U}_1,\textbf{S}_1,\textbf{V}_1]\leftarrow tSVD(\hat{\textbf{G}}_1\hat{\textbf{G}}_2,\delta_{rel.}=\delta_1)$;\\
    $\quad R_1\leftarrow shape(\textbf{S}_1, 1)$;$\hspace{0.1cm}\pmb{\Delta}_1\sim\mathcal{U}([\delta,1])$;\\
    $\quad\hat{\mathcal{G}}_1\leftarrow reshape(\textbf{U}_1\pmb{\Delta}^{-1}_1,[R_0,I_1,R_1])$;\\
    $\quad\hat{\mathcal{G}}_2\leftarrow reshape(\pmb{\Delta}_1\textbf{S}_1\textbf{V}^T_1,[R_1,I_2,R_2])$\;
    \caption{Proposed randomized Tensor Ring-Singular Value Decomposition (rTR-SVD) based on~\cite{zhao2016tensor}.}
    \label{alg:rand_tr_svd}
\end{algorithm}

\emph{Privacy and Correctness.} The correctness of secret sharing based on randomized TN formats is obvious; tensor representations are compressible if the data admits low-rank structure. The proposed randomized tensor decomposition algorithms simply split the complex structural information in big data randomly into different tensor cores or sub-blocks. The sensitivity of SVD decomposition subject to small perturbations is well-known for complex correlation structure, i.e., when the singular values are closely separated~\cite{stewart1998perturbation,liu2008first}. Moreover, the proposed algorithms randomize TN decompositions by large-but-controlled perturbation that does not affect the reconstruction accuracy. The privacy leakage is limited by the tensor-rank complexity of each index, i.e., index that has sufficiently high rank complexity is privacy-preserving, whereas index that has only zeroes in the tensor cores implies that all the values that correspond to this index in the original tensor are zero. However, this can be easily overcome by padding the original tensor with random noise to increase the complexity before TN decomposition. With sufficiently high tensor-rank complexity, the magnitude, sign, and exact position of non-zero values are not leaked even with collusion by all-except-one servers. To further increase the uncertainty, we randomly permute the mode variables along each tensor dimension and store the random seeds for reconstruction. Random permutations can be performed after (block-wise) TN decomposition to ensure compressibility if the multi-dimensional data is highly-correlated. Each tensor core contains only index-specific information and therefore they are unlinkable in the event of massive data breach. The partition of more sophisticated TN structures into private and shared tensor cores can be done with hierarchical clustering based on pairwise network distance and randomized, dispersed tensor computation that minimize privacy leakage, communication, and computational cost.

\emph{Relationship with Additive Secret-Sharing Scheme.}
The classical additive secret-sharing scheme is defined as $x=\langle x_1\rangle_1+\langle x_2\rangle_2+\ldots$. The conversion from the classical scheme to secret-sharing scheme based on TN format is relatively straightforward, each party decomposes their individual share using the proposed randomized tensor decomposition algorithms and send to other parties the corresponding tensor cores. All parties perform an addition operations using their corresponding tensor cores based on tensor multilinear operations. The conversion from TN format to the additive secret-sharing scheme can be done by all-except-one parties generate randomized TN from randomly-generated share, distribute the generated tensor cores to the corresponding party and update all the tensor cores using distributed tensor operations, all-except-one parties pass their updated tensor cores to the remaining one (that didn't generate randomized tensor cores before) to generate his secret share. Future work may consider how to prevent malicious servers from corrupting tensor network computing protocols.

\subsection{Big Data Dispersed Storage, Sharing, and Communication}

Encryption is complicated in terms of key management for big data distributed applications, encryption requires centralized management by a trusted authority to authenticate, authorize, and revoke access to prevent potential key leakage that may lead to massive data breach. Our proposal combines the secret-sharing scheme based on distributed TNs and metadata privacy to seamlessly secure big data storage, communication, and sharing. Distributed trust can be achieved by decentralizing the fragments / metadata encryption and access control mechanisms. Furthermore, the metadata management is flexible such that it can be done in a centralized or decentralized manner by using enterprise management systems, or in a distributed manner on the individual user's side. Any software applications can reconstruct the original data if granted access to the metadata information and shredded fragments. Data integrity can be ensured by cryptographic hashing; whereas data availability can be gueranteed by integrating in Hadoop Distributed File System (HDFS). The advantages of distributed TN representations for secure data storage / sharing include privacy protection, compression, granular access control, updatability, and compressed computation.

Metadata serves as the logical ``map'' for users to navigate through the information and data; metadata also helps auditors to carry out system review and post-breach damage assessment. After decomposing big data and distribute each tensor core or sub-block to multiple storage locations using our proposed randomized tensor decomposition algorithms, the master metadata files are updated with the locations and anonymized filenames of each tensor blocks, tensor structure, cryptographic hashes, random seeds used to permute the mode variables, and users' access permission; the storage locations and filenames of the tensor fragments can be routinely renewed to enhance the data privacy protection. The master metadata files can be further encrypted and password-protected on the users' side. The metadata of each tensor core stored on the distributed storage locations contains only the anonymized filename and location such that they are unlinkable in the event of massive data breach; data encryption and access control based on role management policy can be implemented in a decentralized manner to protect the tensor cores. The system architecture and metadata organization is beyond the scope of this work but will be considered in future to take account of the various application scenarios, system performance, and requirements for different big data applications.

\subsection{Big Data Dispersed Computation}

Tensor network (TN) naturally supports distributed / dispersed computation using the smaller, interconnected tensor cores / blocks after big data decomposition~\cite{cichocki2016tensor,cichocki2017tensor,lee2018fundamental}. Some basic arithmetic operations in Tucker format are derived in~\cite{lee2018fundamental}. Let 
\begin{equation}
\begin{split}
&\mathcal{A}\hspace{0.1cm}=\hspace{0.1cm}[\![\:\mathcal{G}_A;\:\textbf{A}^{(1)},\:\textbf{A}^{(2)},\ldots,\:\textbf{A}^{(N)}\:]\!]\\
&\mathcal{B}\hspace{0.1cm}=\hspace{0.1cm}[\![\:\mathcal{G}_B;\:\textbf{B}^{(1)},\:\textbf{B}^{(2)},\ldots,\:\textbf{B}^{(N)}\:]\!]
\end{split}
\end{equation}
where $\mathcal{G}_{L}$, $L\in\{A,B\}$ and $\textbf{A}^{(k)} / \textbf{B}^{(k)}$, $k\in\{1,2,
\ldots,N\}$ correspond to the Tucker core tensors and factor matrices respectively,
\begin{equation}
\begin{split}
&(a)\:\mathcal{A}+\mathcal{B}=[\![\:\mathcal{G}_A\oplus\mathcal{G}_B\:;\textbf{A}^{(1)}\boxplus\textbf{B}^{(1)},\ldots,\:\textbf{A}^{(N)}\boxplus\textbf{B}^{(N)}\:]\!]\\
&(b)\:\mathcal{A}\oplus\mathcal{B}=[\![\:\mathcal{G}_A\oplus\mathcal{G}_B\:;\textbf{A}^{(1)}\oplus\textbf{B}^{(1)},\ldots,\:\textbf{A}^{(N)}\oplus\textbf{B}^{(N)}\:]\!]\\
&(c)\:\mathcal{A}\circledast\mathcal{B}=[\![\:\mathcal{G}_A\otimes\mathcal{G}_B\:;\textbf{A}^{(1)}\boxtimes\textbf{B}^{(1)},\ldots,\:\textbf{A}^{(N)}\boxtimes\textbf{B}^{(N)}\:]\!]\\
&(d)\:\mathcal{A}\otimes\mathcal{B}=[\![\:\mathcal{G}_A\otimes\mathcal{G}_B\:;\textbf{A}^{(1)}\otimes\textbf{B}^{(1)},\ldots,\:\textbf{A}^{(N)}\otimes\textbf{B}^{(N)}\:]\!]
\end{split}    
\end{equation}
The tensor operations expressed with the symbols $\oplus$, $\boxplus$, $\circledast$, $\otimes$, and $\boxtimes$ refer to the direct sum, partial direct sum, Hadamard product, Kronecker product, and partial Kronecker product respectively, the formal definitions can be found in~\cite{lee2018fundamental}. Linear algebra operations for all tensor formats can be derived using the following rules~\cite{dolgov2013two}:
\begin{itemize}
    \item separable components are added, or multiplied independently in each tensor core for all variables,
    \item all rank sums are added in linear operations, and multiplied in bilinear operations.
\end{itemize}
During iterative computations, the tensor rank grows quickly especially with the multiplications. Hence, another important operation called rank truncation should be provided with the tensor format.

\begin{figure}
\centering
\includegraphics[width=0.9\linewidth]{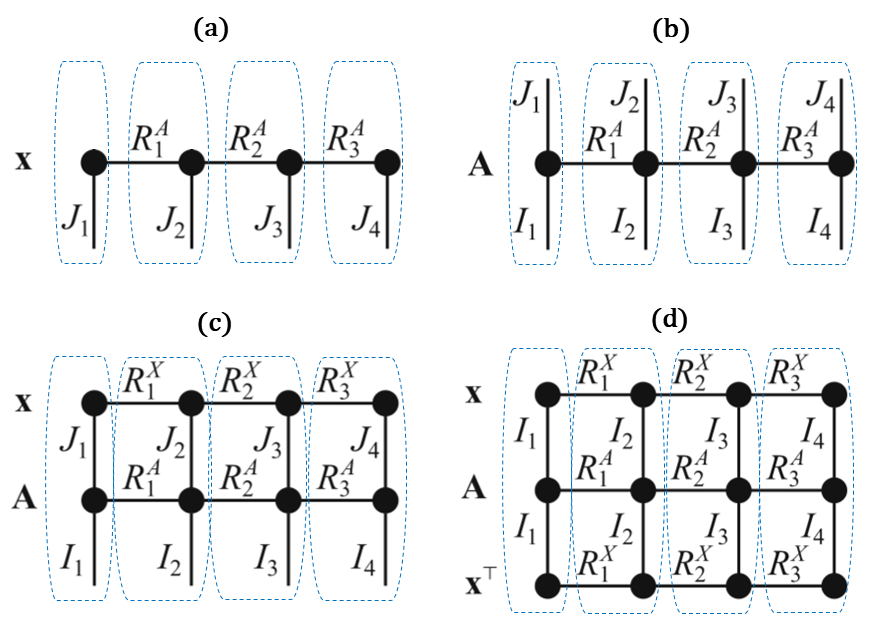}
\caption{Tensor network diagrams of (a) a vector, $\textbf{x}\in\mathbb{R}^{I_1I_2I_3I_4}$ in vector TT format, (b) a matrix, $\textbf{A}\in \mathbb{R}^{I_1I_2I_3I_4\times J_1J_2J_3J_4}$ in matrix TT format, (b) matrix-by-vector multiplication $y=\textbf{A}\textbf{x}$, (c) quadratic form, $\textbf{x}^T\textbf{A}\textbf{x}$ with $I_n=J_n$~\cite{lee2018fundamental}. The dashed / dotted blue boxes show each of the tensor blocks and operations that can be performed in multi-party computation setting.}
\label{fig:tn_net_ops}
\end{figure}

TT format and its variants support wide range of multilinear operations such as addition, multiplication, matrix-by-matrix/vector multiplication, direct sum, Hadamard, Kronecker, and inner product~\cite{lee2018fundamental,cichocki2016tensor,cichocki2017tensor}. As shown in Figure~\ref{fig:tn_net_ops}, multilinear operations in TT format can be performed naturally in dispersed (and compressed) manner, making it well-suited for big data processing and scientific computing. TT-rank grows with every multilinear operations and quickly become computationally prohibitive, the TT-rounding (or recompression)~\cite{oseledets2011tensor} procedure can be implemented to reduce the TT-ranks by first orthogonalizing the tensor cores using QR decomposition and then compress using SVD decomposition, all performed in TT format. The randomized TT-SVD algorithm proposed in Algorithm~\ref{alg:rand_tt_svd} can be easily adapted to the second step of TT-rounding procedure. Algorithm~\ref{alg:rand_tt_round} shows an example of randomized rTT-rounding algorithm for an $N^{th}$-order tensor. To compute non-linear functions, TT cross-approximation can be used~\cite{oseledets2010tt}. The idea of tensor cross or pseudo-skeleton approximation is to sample from the TN, reconstruct and compute arbitrary functions from the sample points, decompose the sample updates and update the original TN accordingly, but how to ensure the privacy preservation of tensor cross approximation is still a question remains.

\begin{algorithm}
    \SetAlgoLined
    \SetKwInOut{Input}{Input}
    \SetKwInOut{Output}{Output}
    \SetKwInOut{Return}{Return}
    \Input{TT cores of an $N^{th}$-order tensor stored on servers, $\mathcal{A}=\langle \textbf{G}_1\rangle_1 \langle \mathcal{G}_2\rangle_{2} \cdots\langle\mathcal{G}_{N-1}\rangle_{N-1}\langle \textbf{G}_N\rangle_N$ and $\epsilon$.}
    \Output{Updated TT cores  $\hat{\mathcal{A}}=\langle\hat{\textbf{G}}_1\rangle_1\langle\hat{\mathcal{G}}_2\rangle_{2}\cdots\langle\hat{\textbf{G}}_N\rangle_N$ such that $\lVert \mathcal{A}-\hat{\mathcal{A}} \rVert_F\leqslant\epsilon\lVert \mathcal{A}\rVert_F$.}
    Initialization: TT-rank $\langle R_0\rangle_1=1$; $\langle R_N\rangle_N=1$;\\
    $\hspace{1.7cm}$ Perturbation threshold $\delta$;\\ 
    $\hspace{1.7cm}$ Truncation parameter $\delta_\epsilon = \frac{\epsilon}{\sqrt{N-1}}$\;
    \textbf{Right-to-left QR orthogonalization:}\\
    $[\langle I_1\rangle_1,\langle R_1\rangle_1]\leftarrow shape(\langle \hat{\textbf{G}}_1\rangle_1)$;\\
    $[\langle R_{N-1}\rangle_N,\langle I_N\rangle_N]\leftarrow shape(\langle \hat{\textbf{G}}_N\rangle_N)$;\\
    $\langle \hat{\mathcal{G}}_1\rangle_1\leftarrow reshape(\langle\hat{\textbf{G}}_1\rangle_1,[R_0,I_1,R_1])$;\\
    $\langle \hat{\mathcal{G}}_N\rangle_N\leftarrow reshape(\langle\hat{\textbf{G}}_N\rangle_N,[R_{N-1},I_N,R_N])$;\\
    \For{$k=N$ to $2$}{
        $[\langle R_{k-1}\rangle_k, \langle I_k\rangle_k, \langle R_k\rangle_k]\leftarrow shape(\langle \mathcal{G}_k\rangle_k)$;\\
        $\langle \textbf{G}_k\rangle_k\leftarrow reshape(\langle \mathcal{G}_k\rangle_k, [R_{k-1}, I_kR_k])$;\\
        QR decomposition: $\left[\langle \hat{\textbf{Q}}_k\rangle_k,\langle \hat{\textbf{R}}_k\rangle_k\right]\leftarrow QR(\langle \textbf{G}_k\rangle_k)$;\\
        $\langle \mathcal{G}_k\rangle_k\leftarrow reshape(\langle \hat{\textbf{Q}}_k\rangle_k,[R_{k-1},I_k,R_k])$;\\
        $\langle \mathcal{G}_{k-1}\rangle_{k-1}\leftarrow \langle \mathcal{G}_{k-1}\times_3\hat{\textbf{R}}_k\rangle_{k-1}$;\\
    }
    \textbf{Left-to-right (SVD compress $+$ random disperse):}\\
    $[\langle R_0\rangle_1, \langle I_1\rangle_1, \langle R_1\rangle_1]\leftarrow shape(\langle \mathcal{G}_1\rangle_1)$;\\
    \For{$k=1$ to $N-1$}{
        $\langle \textbf{G}_k\rangle_k\leftarrow reshape(\langle \mathcal{G}_k\rangle_k, [R_{k-1}I_k, R_k])$;\\ $\left[\langle \textbf{U}_k\rangle_k,\langle \textbf{S}_k\rangle_k,\langle \textbf{V}_k\rangle_k\right]\leftarrow tSVD(\langle \textbf{G}_k\rangle_k,\delta_{rel.}=\delta_\epsilon)$;\\
        $\langle R_k\rangle_k\leftarrow shape(\langle \textbf{S}_k\rangle_k, 1)$; $\langle \pmb{\Delta}_k\rangle_k\sim\mathcal{U}(\delta,1)$;\\
        $\langle \hat{\mathcal{G}}_k\rangle_k\leftarrow reshape(\langle\textbf{U}_k\pmb{\Delta}^{-1}_k\rangle_k,[R_{k-1},I_k,R_k])$;\\
        $\langle\hat{\mathcal{G}}_{k+1}\rangle_{k+1}\leftarrow\langle\hat{\mathcal{G}}_{k+1}\times_1\pmb{\Delta}_{k}\textbf{S}_{k}\textbf{V}^T_{k}\rangle_{k+1}$;\\
    }
    $\langle \hat{\textbf{G}}_1\rangle_1\leftarrow reshape(\langle\hat{\mathcal{G}}_1\rangle_1,[I_1,R_1])$;\\
    $\langle \hat{\textbf{G}}_N\rangle_N\leftarrow reshape(\langle\hat{\mathcal{G}}_N\rangle_N,[R_{N-1},I_N])$;\\
    \caption{Proposed randomized TT-rounding (rTT-rounding) based on~\cite{oseledets2011tensor} to reduce the size of TT-cores.}
    \label{alg:rand_tt_round}
\end{algorithm}

Tensor network computing naturally supports a number of multilinear operations in floating- / fixed-point representations with minimal data pre-processing, unlike classical SMPC schemes that only support limited secure operations (e.g., addition and multiplication) and has to be pre-processed every time to carry out different operations. Therefore, SMPC generally requires many rounds of communication between the servers in order to compute complex functions. With TN representations, multilinear operations can be done in compressed and dispersed manner without the need to reconstruct the original tensor, this is the major advantage of tensor computation in overcoming the curse of dimensionality for large-scale optimization problems. Tensor multilinear operations generally require only computation on each tensor core, but some tensor computation schemes require communication between servers like the TT-rounding scheme mentioned before and the famous Density Matrix Renormalization Group (DMRG) scheme~\cite{schollwock2011density,schollwock2005density}. Unlike SMPC schemes, the communication is mainly tensor cores instead of the secret shares of original tensor, which are generally much smaller in size. However, dispersed tensor computing leaks more information than SMPC schemes during communication, one way to overcome this is to continually ingest fresh entropy from complex data when performing dispersed tensor computation.

\section{Experiments}
\label{sec:exp}

\emph{Experimental Setup.} The experiments are carried out using a workstation with 64-bit Intel\textregistered\hspace{0.01cm} Xeon\textregistered\hspace{0.01cm} W-2123 CPU 3.60GHz, 16.0GB RAM. Privacy metrics such as Pearson's correlation coefficient, histogram analysis, and normalized mutual information are used to measure the privacy leakage of the proposed randomized TN decompositions. Further comparisons are made between the original and the proposed randomized TN decompositions in terms of the computational speed, compression ratio, and distortion analysis of the reconstructed data from TN compression. For image data, the distortion as a result of the TN compression can be measured by the normalized $\mathit{L}_2$-dissimilarity, which is defined by
\begin{equation}
\frac{1}{N'}\sum^{N'}_{n=1}\frac{||\mathbf{x}_n - \mathbf{x}'_n||_2}{||\mathbf{x}_n||_2}
\end{equation}
where $\mathbf{x}_n$, $n\in\{1,2,\ldots,N'\}$ are the set of original images and $\mathbf{x}'_n$, $n\in\{1,2,\ldots,N'\}$ are the set of reconstructed images after TN compression, $||\cdot||$ is the Euclidean norm. Here, we study the proposed rTT-SVD, rTR-SVD, and rTD algorithms only because rHT is based on recursive rTD, therefore showing rTD is privacy-preserving implies that rHT is also privacy-preserving for larger-scale tensor. The perturbation factor $\delta$ for randomized TN is set as $0.05$ for all the experiments, hence the diagonal perturbation matrix $\pmb{\Delta}$ falls within the range $[0.05, 1]$ uniformly.

\emph{Datasets.} Table~\ref{tab:dataset} tabulates all the datasets' sample size and mode size used in this study. Experiments are carried out on 1D, 2D, and 3D biometric datasets to investigate thoroughly the proposed randomized TN algorithms across different data dimensions for privacy preservation. In general, vector and matrix data are reshaped into higher-order tensor before TN decomposition. The gait sensor database is recorded using smartphone's inertial sensors, the sampling frequency is 100Hz and the total walking distance is 640 meters per session~\cite{vajdi2019human}. The training images for real and fake face detection are provided by the Computational Intelligence and Photography Lab, Department of Computer Science, Yonsei University on Kaggle online data-sharing platform; only the real facial images are used in the experiments. The RGB channels of a facial image have very high spectral correlation, therefore the channels are stacked in 3D for tensor decomposition. Yale face database contains the GIF images of 15 human subjects, each with 11 different facial expressions or configurations~\cite{belhumeur1997eigenfaces}. Finally, we also generate a 3D super-diagonal tensor with ones on the $(i,i,i)$ entries for our investigation studies. 

\begin{table}[ht]
\begin{center}
\caption{Datasets used in the experimental studies.}
\label{tab:dataset}
\begin{tabular}{ |c|c|c| } 
 \hline
 \textbf{Dataset}&\textbf{Subjects}&\textbf{Mode Size} \\\hline 
 Human Gait (walking)&93&58 Features\\\hline
 Real \& Fake Face Images&$\sim$1000&600$\times$600$\times$3\\\hline
 Yale Face Database&15&320$\times$243$\times$11\\\hline
 Super-diagonal Tensor&N/A&10$\times$10$\times$10\\\hline
\end{tabular}
\end{center}
\end{table}

\begin{figure}
\centering
\includegraphics[width=1\linewidth]{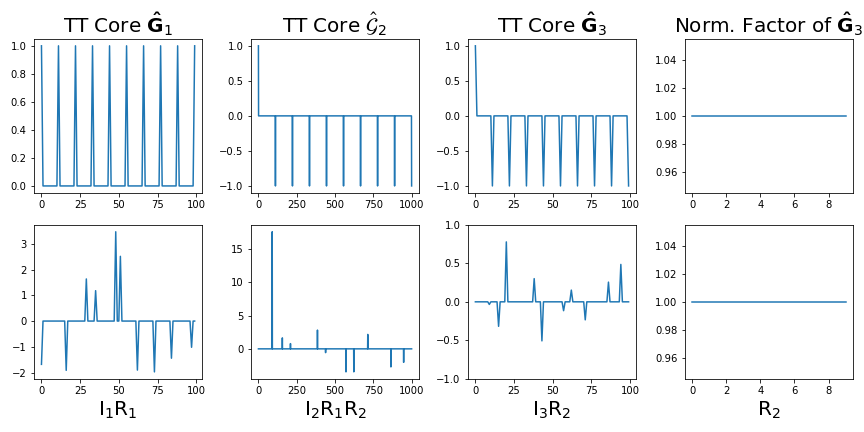}
\caption{TT decomposition of super-diagonal tensor using TT-SVD (top) and rTT-SVD (bottom) proposed in Alg.~\ref{alg:rand_tt_svd}.}
\label{fig:tt_superdiagonal}
\end{figure}

\begin{figure}
\centering
\includegraphics[width=1\linewidth]{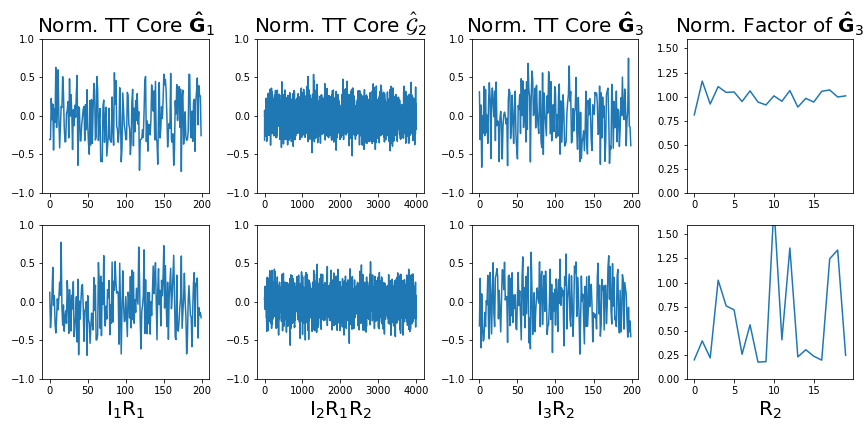}
\caption{TT decomposition of super-diagonal tensor padded with noise using TT-SVD (top) and rTT-SVD (bottom).}
\label{fig:tt_superdiagonal_pad}
\end{figure}

\begin{figure}
\centering
\includegraphics[width=1\linewidth]{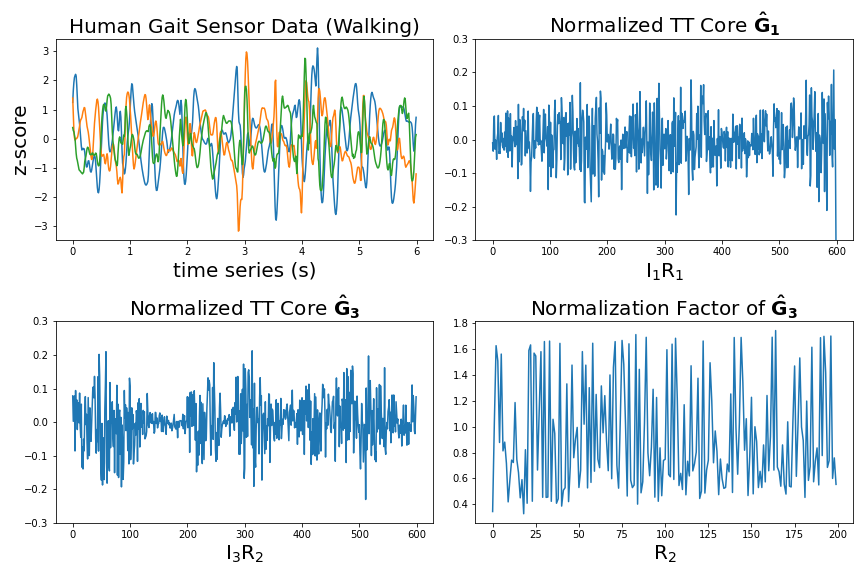}
\caption{Top left: the time series of human gait sensor data in walking mode. Other subplots show the data's TT decomposition. Top right and bottom left: normalized TT core $\mathbf{\hat{G}_1}$ and $\mathbf{\hat{G}_3}$. Bottom right: Normalization factor of $\mathbf{\hat{G}_3}$.}
\label{fig:tt_1d}
\end{figure}

\emph{Data Complexity for Randomized TN Decompositions.} Figures~\ref{fig:tt_superdiagonal} and~\ref{fig:tt_superdiagonal_pad} show the effect on the TT decomposition before and after padding noisy data to a relatively simple (full-rank) super-diagonal tensor, both approaches reproduce the same super-diagonal tensor after reconstruction. However, naive padding with noise usually results in high TN computation and storage cost due to the higher rank-complexity, whereas our proposed randomized TN algorithms simply make use of the complex correlation structure commonly found in big data to generate highly-randomized tensor blocks. Figure~\ref{fig:tt_1d} shows the randomized TT decomposition of human gait sensor data. To preserve important dataset features during the TN compression, each of the attributes is standardized to zero mean and variance equals to one, i.e., the z-score.

\begin{figure}
\centering
\includegraphics[width=1\linewidth]{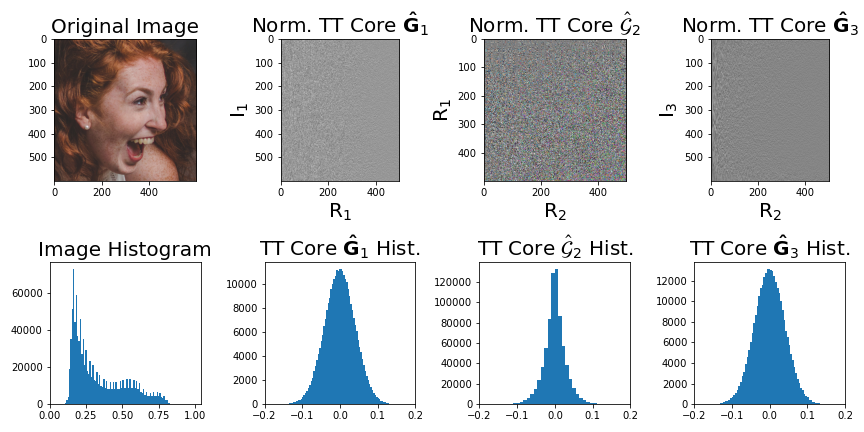}
\caption{Histogram analysis of the TT decomposition of a facial image. The normalized TT cores are either Gaussian or Laplacian distributed, which are usually different from the image histogram distribution.}
\label{fig:tt_frag}
\end{figure}

\begin{figure}
\centering
\includegraphics[width=1\linewidth]{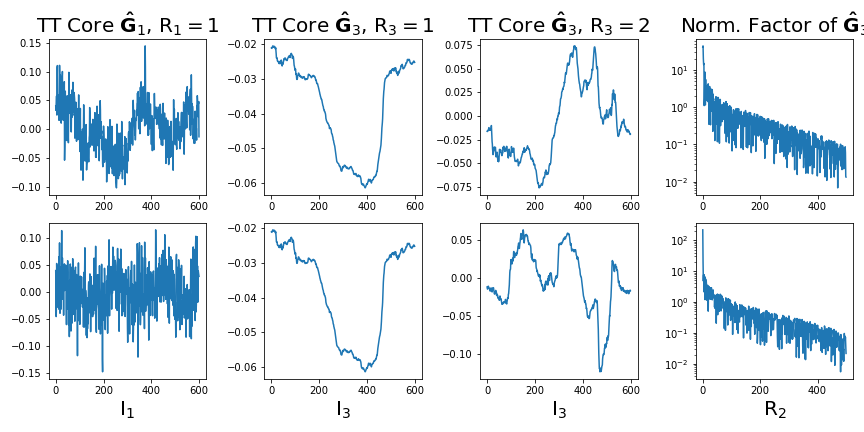}
\caption{Normalized TT cores produced from two randomized rTT-SVD decompositions of a facial image using Algorithm~\ref{alg:rand_tt_svd} (top and bottom rows). Correlation structure that contributes higher variability (i.e., lower rank) is much harder to perturb and the normalization factor in the last TT core is mostly preserved in the randomized decomposition.}
\label{fig:tt_corr}
\end{figure}

\begin{figure}
\centering
\includegraphics[width=1\linewidth]{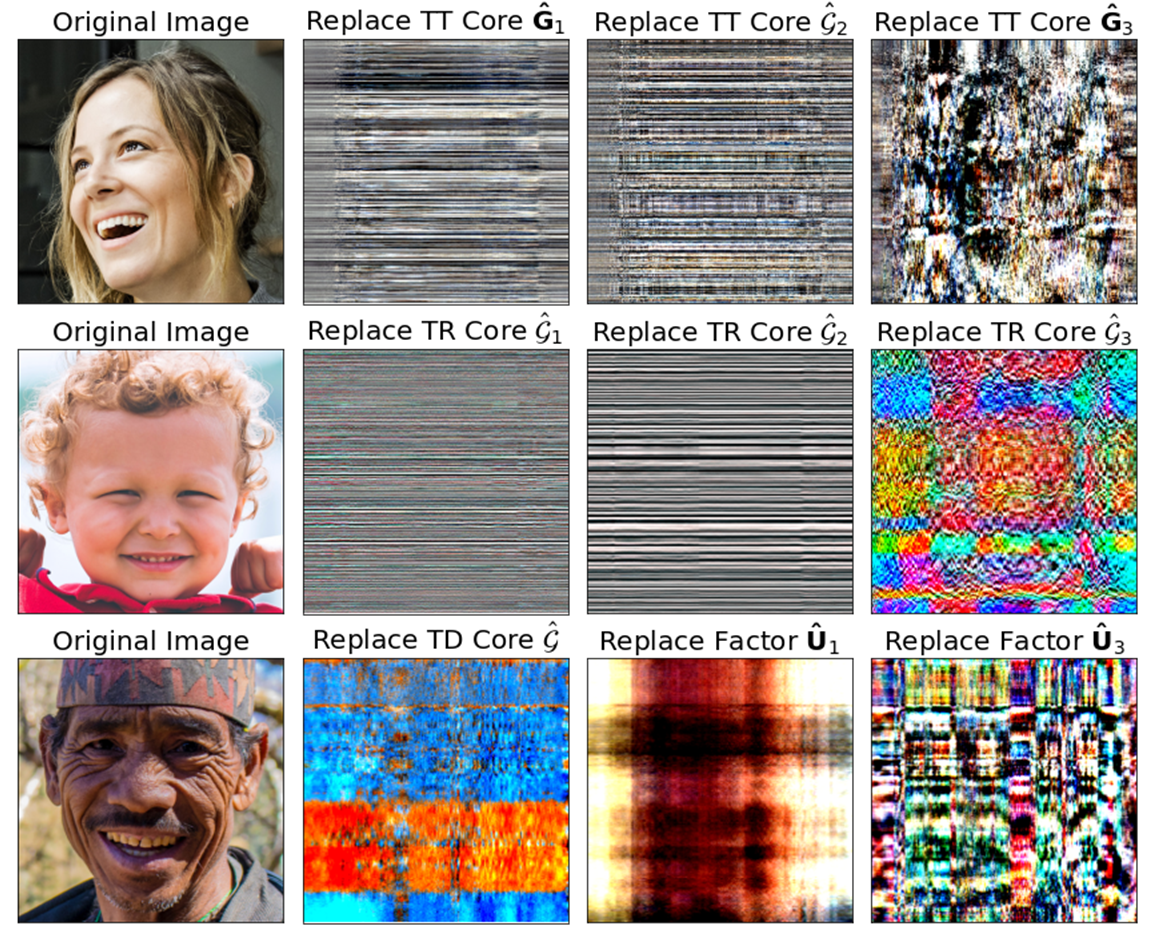}
\caption{Reconstructed images from TN by replacing either a tensor core or factor matrix generated from a randomized TN decomposition process with another. First row corresponds to rTT decomposition, second row is rTR, and third row is rTD respectively.}
\label{fig:mixed_frag}
\end{figure}

\begin{figure}
\centering
\includegraphics[width=1\linewidth]{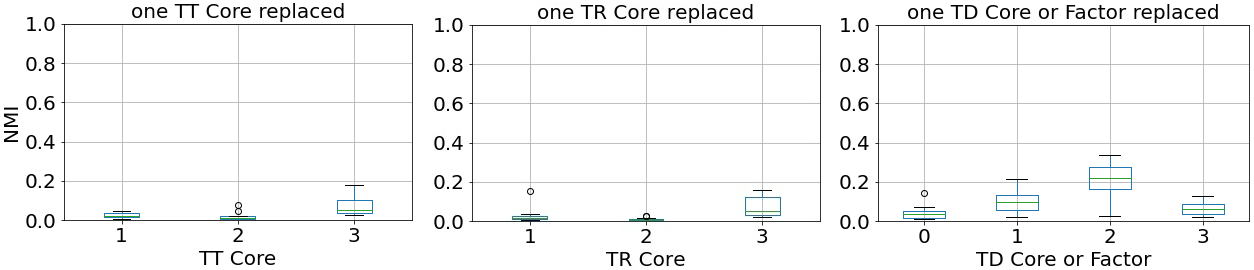}
\caption{Normalized mutual information (NMI) between the original data and the reconstructed data from different randomized TNs with one core or factor replaced.}
\label{fig:tn_nmi}
\end{figure}

\begin{figure}
\centering
\includegraphics[width=1\linewidth]{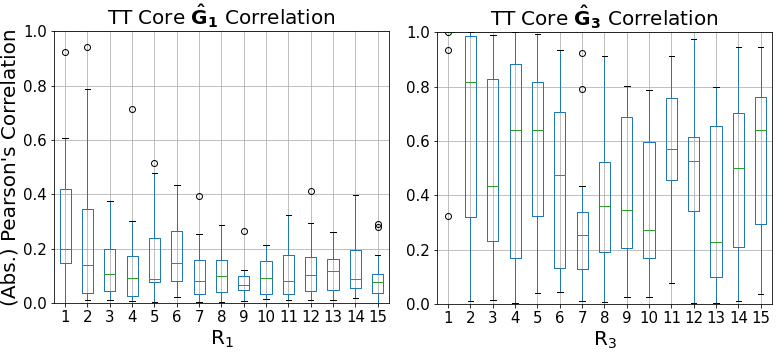}
\caption{Absolute value of Pearson's correlation coefficient for each rank value between the TT cores generated from the original and randomized TT-SVD algorithms. Left: TT core $\mathbf{\hat{G}_1}$. Right: TT core $\mathbf{\hat{G}_3}$. The x-axes refer to $R_1$ and $R_2$ resp.}
\label{fig:tt13_corr}
\end{figure}

\begin{figure}
\centering
\includegraphics[width=1\linewidth]{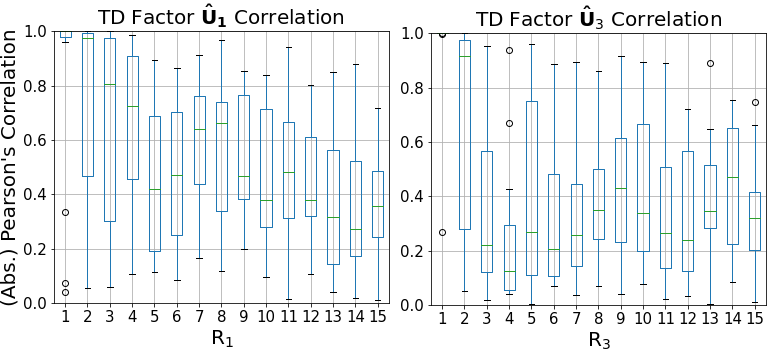}
\caption{Absolute value of Pearson's correlation coefficient for each rank value bet. the Tucker factors generated from the original and randomized TD algorithms. Left: TD factor $\mathbf{\hat{U}_1}$. Right: TD factor $\mathbf{\hat{U}_3}$. The x-axes refer to $R_1$ and $R_3$ resp.}
\label{fig:td_corr}
\end{figure}

\emph{Privacy Leakage Analysis.} Figure~\ref{fig:tt_frag} and~\ref{fig:tt_corr} shows the TT decomposition of a facial image. The shape of the facial image is permuted to $600\times3\times600$ to have a balance shape of TT cores. In general, the histogram of the TN cores and factors are Gaussian or Laplacian distributed, which is very different from the histogram of the original data. Figure~\ref{fig:mixed_frag} and~\ref{fig:tn_nmi} show the reconstructed images from incomplete TN representations and measure the amount of information overlap with original images using normalized mutual information (NMI). The results show that if each of the TN cores or factors are large enough in terms of rank complexity or block size, the privacy leakage is minimal without having a complete TN representations for a data. In this case, the Tucker factor $\mathbf{\hat{U}_2}$ is very small in size and therefore results in the highest NMI. Figures~\ref{fig:tt13_corr} and~\ref{fig:td_corr} show the correlation between the randomized and non-randomized TN cores for particular rank using the Yale Face Database. The correlation is higher for lower rank, this means it is harder to perturb correlation structure that contributes to higher variability within the data. One way to overcome this is to permute the mode variables along each dimension after TN compression to protect the privacy of the distribution of each tensor mode.

\begin{figure}
\centering
\includegraphics[width=1\linewidth]{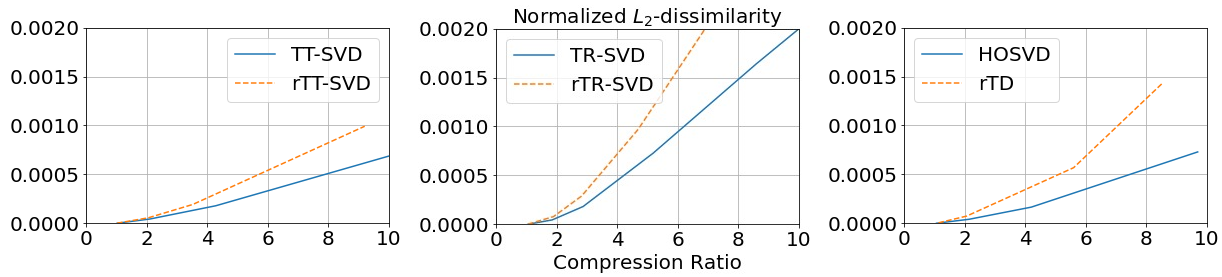}
\caption{Normalized $\mathit{L}_2$-dissimilarity between the original data and the reconstructed data from randomized and non-randomized TN algorithms for diff. compression ratio.}
\label{fig:tn_cr_error}
\end{figure}

\emph{Data Compressibility and Algorithmic Efficiency.} Table~\ref{tab:tn_efficiency} measures the time efficiency of TN decomposition and reconstruction for Real and Fake Facial Image Database. The randomized TN decompositions generally take slightly longer time compared to the non-randomized TN decomposition mainly due to an extra SVD step needed to generate randomized tensor blocks. TR reconstruction is long ($\sim$1 min) because there is a loop in the TN structure. Figure~\ref{fig:tn_cr_error} shows the image distortion analysis under different TN compression ratio for the Yale Face Database. Randomized TN algorithms result in slightly higher distortion in the reconstructed data compared to non-randomized TN algorithms. This is expected because randomized TN algorithms produce sub-optimal decomposition. Randomized TT decomposition generates the lowest distortion especially with high compression ratio compared to randomized rTR and rTD decomposition. TT representation strikes a good balance between privacy preservation, computational, and storage efficiency.
\begin{table}[ht]
\begin{center}
\caption{Comparison bet. the proposed randomized TNs and original algorithms in terms of computational efficiency. The dataset used comes from the Real \& Fake Facial Images Database and the compression ratio is set as $\sim$0.725.}
\label{tab:tn_efficiency}
\begin{tabular}{ |c|c|c| } 
 \hline
 \textbf{Random. TN}&\textbf{Tensor Rank}&\textbf{TN Decompose /}\\
 \textbf{Algorithm}&&\textbf{Reconstruct Time}\\\hline HOSVD&$R_1=R_3=350$, $R_2=3$&0.2794 / 0.0077 s\\\hline
 rTD&$R_1=R_3=350$, $R_2=3$&0.3104 / 0.0081 s\\\hline
 TT-SVD&$R_1=R_2=350$&0.1851 / 0.0054 s\\\hline
 rTT-SVD&$R_1=R_2=350$&0.2817 / 0.0053 s\\\hline
 TR-SVD&$R_{0/1}=R_2=20$, $R_3=45$&0.3563 / 1.1746 s\\\hline
 rTR-SVD&$R_{0/1}=R_2\approx20$, $R_3\approx45$&0.3292 / 1.1150 s\\\hline
\end{tabular}
\end{center}
\end{table}

\section{Discussion}
\label{sec:disc}
Scalability is an important consideration for both the success of big data analytics and widespread adoption of privacy-preserving techniques. We have proposed a simple perturbation technique that can be easily adapted for randomized decomposition of various tensor network structures. The proposed secret-sharing scheme based on dispersed TN representations / computation is very efficient in terms of storage, computational, and communication complexity due to natural support for dispersed tensor computation. Privacy leakage analysis is carried out to verify that the proposed scheme is secured against semi-honest adversary, however, privacy leakage may still happen when performing dispersed tensor operations, which requires further more investigation. One way is to ingest fresh entropy from complex data when performing tensor operations, hence increases the uncertainty of original tensor estimation. Nevertheless, the proposed scheme can be easily combined with existing data-security solutions such as data anonymization, encryption, and secure-enclave technologies to provide layered protection. The potential extension of this work includes various applications of privacy-preserving big data analytics~\cite{cichocki2016tensor,cichocki2017tensor} and large-scale numerical computing~\cite{khoromskij2018tensor,grasedyck2013literature,khoromskij2012tensors}. Another potential direction is extending our proposed secret-sharing scheme for federated machine learning and applying differential privacy to protect the privacy of individual items in the training dataset~\cite{bonawitz2017practical,mcmahan2016communication,shokri2015privacy}.

\clearpage

\bibliographystyle{ACM-Reference-Format}

\end{document}